\documentclass[floatfix,showpacs,superscriptaddress,prl,twocolumn]{revtex4}
\usepackage{amsbsy,amssymb,amsmath,bm}
\usepackage{graphicx,color}
\usepackage{amsfonts}
\usepackage{amssymb}
\usepackage{amsmath}

\input{tcilatex}

\begin{document}

\title{Two distinct ballistic processes in graphene at Dirac point}
\author{M. Lewkowicz}
\affiliation{\textit{Physics Department, Ariel University Center of Samaria, Ariel 40700,
Israel}}
\author{B. Rosenstein }
\email{vortexbar@yahoo.com}
\affiliation{\textit{Electrophysics Department, National Chiao Tung University, Hsinchu
30050,} \textit{Taiwan, R. O. C}.}
\affiliation{\textit{National Center for Theoretical Sciences, Hsinchu 30043,} \textit{%
Taiwan, R. O. C}.}
\affiliation{\textit{Physics Department, Ariel University Center of Samaria, Ariel 40700,
Israel}}
\author{D. Nghiem}
\affiliation{\textit{Electrophysics Department, National Chiao Tung University, Hsinchu
30050,} \textit{Taiwan, R. O. C}.}
\date{\today }

\begin{abstract}
The dynamical approach is applied to ballistic transport in mesoscopic
graphene samples of length $L$ and contact potential $U$. At times shorter
than both relevant time scales, the flight time $t_{L}=L/v_{g}$ ($v_{g}$ -
Fermi velocity) and $t_{U}=\hbar /U$, the major effect of the electric field
is to create electron - hole pairs, i.e. causing interband transitions. In
linear response this leads (for width $W>>L$) to conductivity $\sigma
_{2}=\pi /2$ $e^{2}/h$. On the other hand, at times lager than the two
scales the mechanism and value are different. It is shown that the
conductivity approaches its intraband value, equal to the one obtained
within the Landauer-B\"{u}tticker approach resulting from evanescent waves.
It is equal to $\sigma _{1}=4/\pi $ $e^{2}/h$ for $W>>L$ and $t_{U}<<t_{L}$.
The interband transitions, within linear response, are unimportant in this
limit. Between these extremes there is a crossover behaviour dependent on
the ratio between the two time scales $t_{L}/t_{U}$. At strong electric
fields (beyond linear reponse) the interband process dominates. The electron
- hole mechanism is universal, namely does not depend on geometry (aspect
ratio, topology of boundary conditions, properties of leads), while the
evanescent modes mechanism depends on all of them. On basis of the results
we determine, that while in absorption measurements and in DC transport in
suspended graphene $\sigma _{2}$ was measured, $\sigma _{1}$ would appear in
experiments on small ballistic graphene flakes on substrate.
\end{abstract}

\maketitle

\section{I. Introduction}

Electronic mobility in graphene, especially one suspended on leads, is
extremely large \cite{GeimPRL08}, so that a graphene sheet is one of the
purest electronic systems. The relaxation time of charge carriers due to
scattering off impurities, phonons, ripplons, etc., in suspended graphene
samples of submicron length is so large that the transport is ballistic \cite%
{AndreiNN08,BolotinPRL08}. The ballistic flight time in these samples can be
estimated as 
\begin{equation}
t_{L}=L/v_{g}\,,  \label{t_L}
\end{equation}%
where $v_{g}\simeq 10^{6}m/s$ is the graphene velocity characterizing the
massless "ultrarelativistic" spectrum of graphene near Dirac points, $%
\varepsilon _{k}=v_{g}\left\vert \mathbf{k}\right\vert $, and $L$ is the
length of the sample that can exceed several $\mu m$ \cite{Andrei10}. The
extraordinary physics appears right at the Dirac point at which the density
of states vanishes. In particular, at this point graphene exhibits a quasi -
Ohmic behaviour, $\mathbf{J}=\sigma \mathbf{E}$, even in the purely
ballistic regime.

Determination of the value of the minimal DC conductivity at Dirac point in
the limit of zero temperature has undergone a period of experimental and
theoretical uncertainty. Early on its value, $\sigma \simeq 4e^{2}/h,$ was
measured in graphene on substrate \cite{Novoselov05}, yet consequently it
was shown in experiments on suspended samples \cite{AndreiNN08} that the
zero temperature limit was not achieved, and that, in fact, these early
samples had too many charged "puddles", so that they represented an average
around the neutrality point. The value in early suspended samples \cite%
{AndreiNN08} was half of that and most recently settled around third of this
value in best samples at $2K$ temperature \cite{Andrei10}. Theoretically
several different values for the DC conductivity appeared. The value 
\begin{equation}
\sigma _{1}=\frac{4}{\pi }\frac{e^{2}}{h}  \label{sig1}
\end{equation}%
had been considered as the "standard" one for several years \cite%
{Castro,GusyninPRB06} and appeared as a zero disorder limit of the
self-consistent harmonic approximation \cite{Ziegler06}. It was derived for
the \textit{infinite} sample and this implies the assumption that the
potential difference $U$ at the contacts between the metallic leads and the
graphene flake is unimportant (this is currently under intensive
experimental \cite{Avouris11} and theoretical \cite%
{DoJPCM10,Schomerus07,Datta09,Khomyakov10} investigation). An alternative
and independent approach to ballistic transport in mesoscopic graphene
samples of \textit{finite} length $L$ \cite{Donneau08} with a large contact
barrier $U$ was pioneered in \cite{Beenakker06} following ideas in \cite%
{Katsenelson06}. They applied the Landauer - B\"{u}ttiker formula for
conductance derived for transport in (quasi) one-dimensional channels.

The value 
\begin{equation}
\sigma _{2}=\frac{\pi }{2}\frac{e^{2}}{h}  \label{sig2}
\end{equation}%
was obtained in the dynamical approach to an infinite sample \cite%
{Lewkowicz09} and is equal to the AC value calculated under the condition $%
\omega >>T/\hbar $ at finite temperatures \cite{Ando02,Varlamov07,MacDonald}%
, while other values like $\sigma _{3}=4\frac{e^{2}}{h}$ also appeared in
the literature \cite{Beneventano09}. The dynamical approach to transport was
applied to the tight binding model of graphene \cite{Lewkowicz09} to resolve
this "regularization ambiguity". The ballistic evolution of the current
density in time after a sudden or gradual switching on of the electric field 
$E$ was evaluated and approaches the large times limit $\sigma _{2}E$. The
physical nature of the quasi - Ohmic "resistivity" without either charge
carriers or\ dissipation in infinite samples (namely neglecting leads or
"reservoir") is as follows \cite{Sachdev08,Lewkowicz09}. The electric field
creates electron - hole excitations in the vicinity of the Dirac points
similar to the Landau - Zener tunneling effect in narrow gap semiconductors 
\cite{Davies} or electron - positron pair creation in Quantum
Electrodynamics \cite{Gitman,Nussinov}. Importantly, in graphene the energy
gap is zero, thus the pair creation is possible at zero temperature and
arbitrary small $\mathbf{E,}$ \textit{even within linear response}. Although
the absolute value of the quasiparticle velocity $v_{g}$ cannot be altered
by the electric field due to the "ultrarelativistic" dispersion relation%
\textbf{, }the orientation of the velocity can be influenced by the applied
field.\textbf{\ }The electric current, $e\mathbf{v}$, proportional to the
projection of the velocity $\mathbf{v}$ onto the direction of the electric
field is increased by the field. These two sources of current, namely
creation of moving charges by the electric field (polarization) and their
reorientation (acceleration) are responsible for the creation of a stable
current \cite{Sachdev08,Lewkowicz09,Dora10}. The result within linear
response is that the current settles very fast, on the microscopic time
scale of $t_{\gamma }=\hbar /\gamma \simeq 0.24$ $fs$ ($\gamma $ being the
hopping energy), on the asymptotic value.

A deeper analysis of the "quasi - Ohmic" graphene system beyond the leading
order in perturbation theory in electric field revealed \cite{Kao10} that on
the time scale 
\begin{equation}
t_{nl}=\sqrt{\frac{\hbar }{eEv_{g}}},  \label{t_nl}
\end{equation}%
the linear response breaks down due to intensive Landau-Zener-Schwinger's
(LZS) pair creation and leads to a linear increase with time\cite%
{Rosenstein10}. At times larger than $t_{nl}$ the result is consistent with
the WKB approximation \cite{Cohen,Dora10} This is in contrast to dissipative
systems, in which the linear response limit can be taken directly at
infinite time. This perhaps is the origin of the "regularization"
ambiguities in graphene, since large time and small field limits are
different. Recently the WKB approximation to the interband or the LZS
transition was extended to the finite samples\cite{Vandecasteele10}. In this
case the ballistic evolution is "truncated" at ballistic time $t_{L}$, Eq.(%
\ref{t_L}). It is (sometimes implicitly) assumed within both the dynamical
approach and the Kubo approach that there is no significant contact barrier $%
U$ between the leads and the graphene flake.

In contrast, the Landauer - B\"{u}ttiker (LB) approach hinges on the
description of the leads in terms of a potential barrier of a certain
non-zero barrier height $U\left( r\right) $ \cite{DoJPCM10}. The first
quantized Weyl equation therefore was considered in ref. \cite{Beenakker06}
in full analogy to the one-particle Schr\"{o}dinger equation to define the T
- matrix and it was found, surprisingly, that quasi - Ohmic behaviour
emerges for large aspect ratios $W/L\rightarrow \infty $ and moreover the
conductivity is precisely $\sigma _{1}$ for rectangular samples. The
approach was extended to barriers of various shapes \cite{Sonin}, boundary
conditions \cite{Schomerus07}, nonrectangular geometries like Corbino disks
using conformal mapping \cite{Rycerz09}. It was established that the
conductance is quite sensitive to the topology of the sample, but rather
insensitive to the potential shape. This approach cannot be extended to the
AC fields. This is an additional motivation to extend the dynamical approach
to include the effects of the barrier potential. The barrier potential
provides an additional time scale%
\begin{equation}
t_{U}=\hbar /U.  \label{t_U}
\end{equation}

It looks like the physical picture behind the LB approach is almost
"orthogonal" to the one of the ultrarelativistic pair creation mentioned
above. In this note we rigorously apply the dynamical approach to study
transport in mesoscopic samples. We demonstrate that the physics behind the
two values of the DC conductivity is quite different despite the fact that
numerically $\sigma _{2}=1.57e^{2}/h$ is just 24\% higher than $\sigma
_{1}=1.27e^{2}/h$ for the stripe geometry (this is quite accidental due to
the nonuniversality of the latter value \cite{Rycerz09}). A recent numerical
simulation of the dynamics of a graphene ribbon \cite{Cini10} demonstrates
that the value of conductance corresponding to $\sigma _{2}$ on a short time
scale crosses over at $t_{L}$ (via a series of strong oscillations) into one
corresponding to the LB conductance in the ribbon (analogous to $\sigma _{1}$%
) at asymptotically long times. These two physical processes governing the
ballistic transport are quite distinct. One is fast and homogeneous: the
interband channel (valence and conduction "cones" of graphene near Dirac
point), namely the electron - hole creation sometimes referred to (usually
beyond linear response) as Landau - Zener tunneling, or, in particle
physics, the Schwinger's pair creation\cite{Schwinger}. It is unique to
graphene (and some other systems with similar band structure like
topological insulators) and has certain surprising features. For example,
this channel of conduction "dries out" or is depleted for any finite sample.
The second mechanism, the intraband transition, despite constituting a
peculiar "relativistic" kind of electron acceleration, is much more common.
It is important for transport only for a sufficiently large contact
potential between the leads and the graphene sample and unlike the interband
channel, is a long time phenomenon.

We use the dynamical approach to determine what process is dominant for the
evolution of the I-V curve of a finite graphene sample directly at the
neutrality point, with the contact barrier taken into account. We explore
this evolution on different time scales $t$, which can be associated with
the frequency $\omega \sim 1/t$ for a periodic pulse or the pulse duration
in the relaxation type experiments \cite{Rusin,Rana}. The physics depends
essentially on the relation of a time scale $t$ with respect to the three
physical time scales $t_{L},t_{U}$ and $t_{nl}$ defined above in Eqs.(\ref%
{t_L},\ref{t_U},\ref{t_nl}). It is demonstrated analytically that for a
finite barrier potential and finite length the infinite time limit coincides
in linear response with a generalization of the LB calculation in \cite%
{Beenakker06}.

We start in Section II from the definition of a model neglecting effects of
the contact potential barrier, namely for $t_{U}>>t_{L},t_{nl}$. The
dynamical approach to ballistic transport in the infinite sample is briefly
outlined (this way $\sigma _{2}$ is obtained). The approach is generalized
to the case of a finite sample and the decay of the interband channel is
demonstrated. In Section III a phenomenological model of the graphene - lead
coupling is specified and certain stationary properties, like the quasi one-
particle T - matrix, are constructed. The LB results are slightly
generalized to the case of an arbitrary barrier height $U$ equal to the
chemical potential of the leads. It will be interpreted as a long time limit
of the intraband contribution within the dynamical approach to the finite
sample formulated in Section IV. The evolution of the current in graphene at
Dirac point for an arbitrary potential barrier is given here (within linear
response) as an integral. Results of the numerical evaluation of the
integral together with the analytically obtained short- and the long time
asymptotes are given in Sections V and VI for the intraband and the
interband contributions, respectively. The results are summarized and
discussed in Section VI.

\section{II. Small contact barrier: dynamical approach to the interband
transition}

\subsection{A. The Hamiltonian for the infinite sample}

The electrons in a constant and homogeneous electric field near the "right
helicity" Dirac point\cite{Castro} are approximately described by the Weyl
Hamiltonian, 
\begin{eqnarray}
\hat{H} &=&\hat{K}+\hat{V};  \label{H} \\
\hat{K} &=&\int d^{2}r\text{ }\hat{\psi}_{r}^{\dagger }K\hat{\psi}_{r};\text{
\ }K=-i\hbar v_{g}\mathbf{\sigma }\cdot \mathbf{\nabla };  \label{Kdef} \\
\hat{V} &=&e\int d^{2}r\text{ }\hat{\psi}_{r}^{\dagger }V\hat{\psi}_{r};%
\text{ \ }V=\frac{\hbar v_{g}}{c}\mathbf{\sigma }\cdot \mathbf{A}+\Phi ;
\label{Vdefin}
\end{eqnarray}%
where $K$ and $V$ are first quantized operators and the annihilation
operator $\hat{\psi}_{r}^{\alpha }$ is a two component $\alpha =1,2$
(pseudo) spinor. $\mathbf{A}$ and $\Phi $ are the vector and the scalar
potentials describing the electric field which is switched on at $t=0,$
oriented along the $y$ axis and, importantly, is coordinate independent. We
employ units in which $\hbar =v_{g}=1$. In momentum basis, $\hat{\psi}_{%
\mathbf{r}}=\frac{1}{\sqrt{WD}}\dsum\limits_{\mathbf{k}}e^{i\mathbf{k}\cdot 
\mathbf{r}}\hat{\psi}_{\mathbf{k}},$ where $D$ is an infrared cutoff and $W$
is the width that also will be treated as large. When the constant electric
field is written in a gauge that respects the translational symmetry, $\Phi
=0$, $\mathbf{A=}\left( 0,-cEt\right) $, the different momenta decouple: 
\begin{equation}
\hat{H}=\sum_{\mathbf{k}}\hat{\psi}_{\mathbf{k}}^{\dagger }\mathbf{\sigma }%
\cdot \left( \mathbf{k+}\frac{e}{c}\mathbf{A}\right) \hat{\psi}_{\mathbf{k}}%
\text{.}  \label{H_matrix}
\end{equation}%
The model is a rather crude idealization of the experimental situation in
several respects. The sample is considered "infinite" and absolutely
homogeneous. This allows a convenient use of the axial gauge invariant under
translations. Finite width (perpendicular to the electric field ) generally
creates no complications and the whole discussion can be repeated for finite 
$W$ and variety of boundary conditions. On the other hand, finite length $L$
breaks the translational invariance and simplicity is lost; this is one of
the subjects of the present paper. The second idealization pertains the
description of leads. It is assumed that the leads are absolutely
unintrusive, namely, one can imagine the Corbino disc geometry or a lead
with no contact potential difference. The more general case, with a
potential barrier, will be treated below.

The spectrum before the electric field is switched on is divided into
positive and negative energy parts describing the valence and conduction
band:%
\begin{eqnarray}
\left( \mathbf{\sigma }\cdot \mathbf{k}\right) u_{\mathbf{k}}
&=&-\varepsilon _{\mathbf{k}}u_{\mathbf{k}};\text{ \ \ \ \ }\left( \mathbf{%
\sigma }\cdot \mathbf{k}\right) v_{\mathbf{k}}=\varepsilon _{\mathbf{k}}v_{%
\mathbf{k}};  \label{uvdef} \\
u_{\mathbf{k}} &=&\frac{1}{\sqrt{2}}%
\begin{pmatrix}
1 \\ 
-z_{\mathbf{k}}%
\end{pmatrix}%
;\text{ \ \ \ \ \ }v_{\mathbf{k}}=\frac{1}{\sqrt{2}}%
\begin{pmatrix}
1 \\ 
z_{\mathbf{k}}%
\end{pmatrix}%
,  \label{uv}
\end{eqnarray}%
where $z_{\mathbf{k}}=\left( k_{x}+ik_{y}\right) /\varepsilon _{\mathbf{k}}$
is a phase and $\varepsilon _{\mathbf{k}}=\left\vert \mathbf{k}\right\vert $%
. Since all the momenta are independent in the Hamiltonian Eq.(\ref{H_matrix}%
) due to use of the gauge in which the electric field is represented via a
homogeneous vector potential, a second quantized state\ is uniquely
characterized by the "first quantized" amplitude, 
\begin{equation}
\psi _{\mathbf{k}}\left( t\right) =%
\begin{pmatrix}
\psi _{\mathbf{k}}^{1}\left( t\right) \\ 
\psi _{\mathbf{k}}^{2}\left( t\right)%
\end{pmatrix}%
\text{,}  \label{spinor}
\end{equation}%
which is a "spinor" in the sublattice space. It obeys the matrix Schr\"{o}%
dinger equation in sublattice space: $i\partial _{t}\psi _{\mathbf{k}}=%
\mathbf{\sigma }\cdot \left( \mathbf{k+}\frac{e}{c}\mathbf{A}\right) \psi _{%
\mathbf{k}}$. The initial condition corresponding to a second quantized
state at zero temperature in which all the negative energy states are
occupied and all the positive energy states are empty is $\psi _{\mathbf{k}%
}\left( t=0\right) =u_{\mathbf{k}}$.

The evolution of the current density, $\mathbf{\hat{J}}=-4e\hat{\psi}%
_{r}^{\dagger }\mathbf{\sigma }\hat{\psi}_{r}$ , of a state in terms of this
amplitude is

\begin{equation}
j_{y}\left( t\right) =-4e\sum_{\mathbf{k:\varepsilon }_{\mathbf{k}}<\mu
}\psi _{\mathbf{k}}^{\dag }\left( t\right) \sigma _{y}\psi _{\mathbf{k}%
}\left( t\right) \text{.}  \label{Jy_def}
\end{equation}%
The factor $4$ is due to spin and valley degeneracies of the Weyl fermions.
To leading order in the DC electric field for $\mu =0$ one obtains \cite%
{Lewkowicz09} $\sigma =\sigma _{2}$, Eq.(\ref{sig2}). The same result is
obtained for any frequency \cite{Varlamov07, MacDonald}. It was calculated
analytically in the tight binding model for arbitrary $E$ (beyond linear
response) in ref. \cite{Rosenstein10}. Corrections to both DC and AC
conductivity were computed in \cite{Kao10} and reveal that the linear
response breaks down at $t_{nl}$ as mentioned above and is perhaps a source
of the "regularization ambiguity" in linear response. The dynamical approach
provides a simple interpretation for the nature of the excitations: the
copious creation of electron - hole pairs or interband transitions. Beyond
linear response these transitions can be treated either within the Landau -
Zener (or WKB) approximation \cite{Dora10,Vandecasteele10,Nussinov} or
exactly by using Schwinger's method \cite{Kao10}.

The simple method of calculation used in the above works hinges on the
translational invariance of both the sample and the electric field. However,
as long as the electric field is treated within linear response only, one
may also consider an inhomogeneous bias. It turns out that in graphene at
Dirac point, if there is no potential barrier like the one assumed within
the LB approach (and discussed in detail below), the current density decays
with time even as the local electric field stays constant.

\subsection{B. Linear response: decay of the current in a finite sample with
a finite range of the electric field}

To model the bias voltage we assume that electric field is homogeneous in
the segment $-L/2<y<L/2$, and therefore can be described by a scalar
potential,

\begin{equation}
\Phi \left( y\right) =\frac{V_{0}}{2}\left\{ 
\begin{array}{c}
1\text{ for }y<-L/2 \\ 
-2y/L\text{ for }-L/2<y<L/2 \\ 
-1\text{ for }L/2<y%
\end{array}%
\right. \text{,}  \label{V}
\end{equation}%
(and $\mathbf{A}=0$), see the dashed line in Fig.1. This assumption holds
even in high current experimental situations like the one described in ref. 
\cite{Vandecasteele10}. The current, to leading order in perturbation $\hat{V%
}$, Eq.(\ref{Vdefin}) for Dirac point $U_{gate}=0$ is:%
\begin{eqnarray}
&&I_{y}\left( t\right)  \label{Inobarrier} \\
&=&-4W\sum\limits_{\mathbf{l,p}}\frac{1-e^{-i\left( \varepsilon _{\mathbf{p}%
}+\varepsilon _{\mathbf{l}}\right) t}}{\varepsilon _{\mathbf{p}}+\varepsilon
_{\mathbf{l}}}\left\langle u_{\mathbf{l}}\left\vert \hat{V}\right\vert v_{%
\mathbf{p}}\right\rangle \left\langle v_{\mathbf{p}}\left\vert \hat{J}%
_{y}\right\vert u_{\mathbf{l}}\right\rangle +cc\text{.}  \notag
\end{eqnarray}%
As explained in detail in Sec. IV of ref. \cite{Kao10}, the current within
the Weyl model has an ultraviolet divergence that should be removed in a
chirally invariant manner. Since the present case is not different in this
respect from that of the infinite range electric field, the details are
omitted. After some algebra, the conductance (for large $W$ so that
continuum momentum can be used) takes the form%
\begin{widetext}%

\begin{equation}
G\left( t\right) =\frac{We^{2}}{\pi ^{3}L}\int_{k=-\infty }^{\infty
}\int_{p,l=0}^{\infty }\frac{\sin \left[ \left( p-l\right) L/2\right] \left(
l\varepsilon _{kp}-p\varepsilon _{kl}\right) }{\left( p-l\right)
^{2}\varepsilon _{kl}\varepsilon _{kp}}\frac{\sin \left[ t\left( \varepsilon
_{kp}+\varepsilon _{kl}\right) \right] }{\varepsilon _{kp}+\varepsilon _{kl}}%
\text{,}  \label{g(t)nobar}
\end{equation}

\end{widetext}%
where $k=l_{x}=p_{x},l=l_{y},p=p_{y}$. In terms of dimensionless variables $%
\overline{\epsilon }=\left( \varepsilon _{kp}+\varepsilon _{kl}\right) L$, $%
\delta =L\left( \varepsilon _{kp}-\varepsilon _{kl}\right) /\overline{%
\epsilon }$, $\Delta =L\left( p-l\right) /\overline{\epsilon }$, one obtains
the (effective) scaled conductivity $\sigma \left( \overline{t}\right)
\equiv G\left( t\right) \frac{L}{W}$ as function of time in units of the
flight time $\overline{t}=t/t_{L}$:%
\begin{equation}
\small%
\sigma \left( \overline{t}\right) =-\frac{4e^{2}}{\pi ^{3}}\int_{\overline{%
\epsilon }=0}^{\infty }\frac{\sin \left( \overline{t}\overline{\epsilon }%
\right) }{\overline{\epsilon }}\int_{\delta =0}^{1}\int_{\Delta =\delta }^{1}%
\frac{1}{\Delta ^{3}}\sqrt{\frac{\Delta ^{2}-\delta ^{2}}{1-\Delta ^{2}}}%
\sin \left( \frac{\overline{\epsilon }\Delta }{2}\right) \text{.}
\label{G(t)_small}
\end{equation}%
This function is given as the red line Fig.2 and has the following
behaviour. Before $\overline{t}=1/2$, $\sigma \left( \overline{t}\right)
=e^{2}/4$; therefore in physical units one recovers the "dynamical" value $%
\sigma _{2}=\frac{\pi }{2}\frac{e^{2}}{h}=\frac{e^{2}}{4\hbar }.$This is
just the result of pseudo-relativistic invariance (maximal velocity $v_{g}$)
of the Weyl model. The effect of the finite extent of the electric field has
no time to propagate to the center of the sample where the current is
defined. Then the current drops fast and settles at $t_{L}$ into a power
decrease

\begin{equation}
\sigma \left( \overline{t}\right) =\frac{e^{2}L}{4\pi t}=\sigma _{2}\frac{1}{%
\pi \overline{t}}\text{.}  \label{sig_small}
\end{equation}%
In the whole range $\overline{t}>1/2$ there is an excellent fit for this
function$:$%
\begin{equation}
\sigma \left( \overline{t}\right) =\sigma _{2}\frac{1}{\pi \overline{t}-\pi
/2+1-\sqrt{\overline{t}/2-1/4}}\text{.}  \label{sig_interpolation}
\end{equation}

Until now the linear response approximation was used. Hence, for a finite
range of the electric field (finite distance between the electrodes) a
stationary flow state is only possible beyond linear response. The LZS
tunneling over the band gap is generally a nonperturbative phenomenon. The
linear response is only useful in a limited time, $t<t_{nl},$ due to the
ultrarelativistic spectrum of graphene \cite{Rosenstein10}. The interband
processes beyond $t_{nl}$ are discussed next.

\subsection{C. The LZS conductance beyond linear response}

Since there are two characteristic times beyond linear response, $t_{L}$ and 
$t_{nl}$, three possibilities exist: $t_{L}>>t_{nl}$, $t_{L}<<t_{nl}$ or $%
t_{L}\sim t_{nl}$. In most transport experiments the ratio $t_{L}/t_{nl}=L%
\sqrt{eE/\hbar v_{g}}$ is smaller than $1$. The time scale on which
nonlinear effects become dominant is not always very large; for example, for
experiments not necessarily dedicated to large current transport
measurements \cite{SinghPRB09} in which $E=10^{4}V/m$ and $L=0.3\mu m$,
nonlinearity sets in at $t_{nl}=0.3ps,$ which is of order of the ballistic
time. Moreover, graphene flakes under large fields of order $2\cdot
10^{6}V/m $ have been recently studied in specially designed high current
density experiments \cite{Vandecasteele10}. For a sample length $L=2\mu m$
this results in a very large ratio $t_{L}/t_{nl}=100$. Therefore it is of
importance to calculate the conductance for arbitrary $t_{L}/t_{nl}.$ We
start from large fields for which the LZS tunneling is the most effective.

\subsubsection{$t_{L}>>t_{nl}$}

Analytic and numerical solutions of the tight binding model\cite%
{Rosenstein10}, as well as of the Weyl model describing the physics near the
Dirac point demonstrated\cite{Kao10,Dora10} that at $t_{nl}$ the creation of
electron - hole pairs becomes dominant and is well described by an
adaptation of the well - known Schwinger electron - positron pair creation
rate $\frac{d}{dt}N\propto \left( eE\right) ^{3/2}$. The difference with the
original derivation \cite{Schwinger} in the context of particle physics is
that the fermions are 2+1 dimensional and "massless", thus magnifying the
effect. The polarization current is $J\left( t\right) =-2ev_{g}N\left(
t\right) $ and therefore Schwinger's creation rate at asymptotically long
times leads to a linear increase with time: 
\begin{equation}
\sigma \left( t\right) =\sigma _{2}\left( eE\right) ^{1/2}t\text{.}
\label{J_nl}
\end{equation}%
Interestingly this formula is very accurate already at $t=t_{nl}$, see ref. 
\cite{Rosenstein10}.

The physics of pair creation is highly non-perturbative and non-linear in
nature and therefore, instead of the linear response, Schwinger had to use
functional methods to get an exact asymptotic formula. The rate can be
intuitively understood using the much simpler instanton approach originally
proposed in the context of particle physics \cite{Nussinov} (extended later
to low dimensions \cite{Gitman}), that is known in condensed matter physics
as the Landau - Zener tunneling probability \cite%
{Dora10,Santoro,Vandecasteele10}.

The density in the infinite sample, calculated using the simple Landau -
Zener creation rate expression for one of the flavours, is \cite%
{Dora10,Nussinov}:

\begin{equation}
N_{\mathbf{k}}\left( t\right) =\Theta \left( k_{y}\right) \Theta \left(
eEt-k_{y}\right) \exp \left( -\pi k_{x}^{2}/eE\right) \text{,}
\label{LandauZener}
\end{equation}%
where $\Theta $ are the Heaviside functions. One considers the "tunneling"
from the conduction band to the valence band at fixed $k_{x}$. The gap is
given by $2\left\vert k_{x}\right\vert $. Consequently the number of pairs is

\begin{equation}
N\left( t\right) =\frac{4}{\left( 2\pi \right) ^{2}}\int_{\mathbf{k}}N_{%
\mathbf{k}}\left( t\right) =\frac{4}{\pi ^{2}}\left( eE\right) ^{3/2}t\text{,%
}  \label{N}
\end{equation}%
corresponding to the Schwinger's pair creation rate $\frac{d}{dt}N=\frac{4}{%
\pi ^{2}}\left( eE\right) ^{3/2}$. There is a simple relation within the
Weyl model between the rate and the current density, as was shown recently 
\cite{Dora10} (see also \cite{Gitman}): 
\begin{widetext}%

\begin{equation}
J=\frac{2e}{\pi ^{2}}\int_{\mathbf{k}}\left[ \frac{eEt-k_{y}}{\sqrt{%
k_{x}^{2}+\left( eEt-k_{y}\right) ^{2}}}N_{\mathbf{k}}\left( t\right) +\frac{%
\sqrt{k_{x}^{2}+\left( eEt-k_{y}\right) ^{2}}}{eE}\frac{d}{dt}N_{\mathbf{k}%
}\left( t\right) \right]  \label{JcJp}
\end{equation}
\end{widetext}
Substituting Eq.(\ref{LandauZener}), one obtains:

\begin{equation}
J\left( t\right) =\frac{2e}{\pi ^{2}}\int_{k_{x}=0}^{\infty }\left( \sqrt{%
k_{x}^{2}+\left( eEt\right) ^{2}}+k_{x}\right) \exp \left( -\frac{\pi
k_{x}^{2}}{eE}\right) .  \label{Jc}
\end{equation}%
At large $t$, one arrives at conductivity, Eq.(\ref{J_nl}).

Adaptation of the instanton approach to finite length sample is quite
cumbersome, however the long time limit is simple, as was shown in ref.\cite%
{Vandecasteele10}. The range of integration over momenta is determined by
the semiclassical condition for tunneling:

\begin{equation}
J\left( t\rightarrow \infty \right) =\frac{2e}{\pi ^{2}}%
\int_{k_{x}=0}^{eEL/2}\int_{k_{y}=0}^{k_{y}^{\max }}\frac{k_{y}}{\varepsilon
_{\mathbf{k}}}\exp \left( -\frac{\pi k_{x}^{2}}{eE}\right) ,  \label{Jsemicl}
\end{equation}%
where $k_{y}^{\max }=\sqrt{\left( eEL\right) ^{2}-k_{x}^{2}}$. The result
for conductivity, 
\begin{eqnarray}
&&\sigma \left( t\rightarrow \infty \right) =  \label{sig_inst} \\
&&\frac{2e^{2}}{\pi ^{3}}\left[ \frac{\pi L}{2}\sqrt{eE}\func{erf}\left( 
\frac{L}{2}\sqrt{\pi eE}\right) +\exp \left( -\frac{\pi eEL^{2}}{4}\right) -1%
\right] ,  \notag
\end{eqnarray}%
is presented as a green line in Fig. 2. For $t_{L}>>t_{nl}$ it is
proportional to the ratio $t_{L}/t_{nl}:$%
\begin{equation}
\sigma \left( t\rightarrow \infty \right) =\frac{e^{2}L}{\pi ^{2}}\sqrt{eE}=%
\frac{4}{\pi ^{2}}\sigma _{2}\frac{t_{L}}{t_{nl}}\text{.}
\label{sig_inst_large}
\end{equation}%
This is larger than $\sigma _{2}$ for $t_{L}/t_{nl}>\pi ^{2}/4\approx 2.5$.
For these fields the slow decrease due to finite extent of electric field, $%
L $, does not materialize and increases monotonically as function of time $t$
(physically representing the pulse period or pulse duration). For small
ratios $t_{L}/t_{nl}$ the situation is different.

\subsubsection{$t_{L}<<t_{nl}$}

In this case of small electric fields one first encounters at $t_{L}$ a
powerwise drop in current density from the short time value $\sigma _{2}$ as
in Eq.(\ref{sig_small}) before the nonlinear effects take over. For large
times $t>t_{nl}$ one can use the same semiclassical LZ method to determine
how the powerwise decrease is halted. The asymptotic value given by Eq.(\ref%
{sig_inst}) is this case simplifies into:

\begin{equation}
\sigma \left( t\rightarrow \infty \right) =\frac{2e^{2}}{\pi ^{2}}\left(
eE\right) L^{2}=\frac{8}{\pi ^{2}}\sigma _{2}\left( \frac{t_{L}}{t_{nl}}%
\right) ^{2}\text{.}  \label{sig_inst_small}
\end{equation}%
This intercept with the slow decreasing current, Eq.(\ref{sig_small}), see
Fig.2, occurs at very large time $t=\frac{\pi }{8}\frac{t_{nl}^{2}}{t_{L}}$.
When $t_{L}$ is of the same order as $t_{nl}$, the effective conductivity is
approximately $\sigma _{2}$, see values close to $t_{L}/t_{nl}=\pi ^{2}/4$.

Till now we neglected the influence of a potential barrier between the leads
and graphene on the ballistic transport.

\begin{figure}[ptb]\begin{center}
\includegraphics[
natheight=1.9519in, natwidth=3.2085in, height=1.9908in, width=3.2534in]
{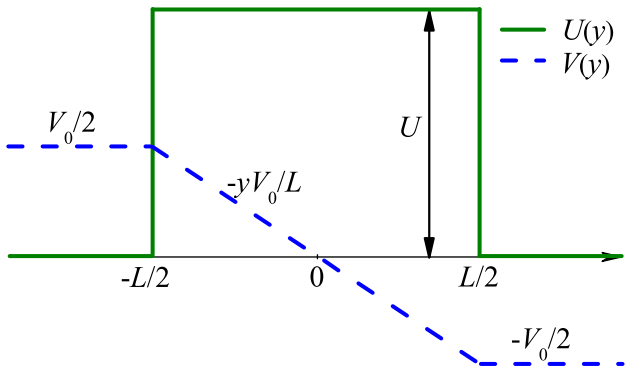}\caption{Potential barrier
U(y) (green line) describes contacts of the leads with the sample, while the
bias electric potential V(y) (blue dashed line) describes the applied
constant electric field (from -L/2 to L/2).}
\end{center}\end{figure}%

\section{III. Contact barrier: stationary properties and the first quantized
Landauer - B\"{u}ttiker approach}

\subsection{A. Phenomenological description of contacts. Symmetry of the
Hamiltonian.}

One models the effect of coupling to leads by a finite (and sometimes very
large \cite{Beenakker06}) potential energy barrier (that should in principle
be found self consistently \cite{DoJPCM10}). The simplest model is the
square barrier, see Fig.1,

\begin{equation}
U\left( y\right) =\left\{ 
\begin{array}{c}
0\text{ for }y>L/2\text{ or }y<-L/2 \\ 
U>0\text{ for}-L/2<y<L/2%
\end{array}%
\right. \text{.}  \label{U}
\end{equation}%
The derivation of the model from a microscopic Hamiltonian is discussed in
several works, see for example \cite{DoJPCM10} and it was found to describe
a typical transport experiment quite well.

The second quantized Hamiltonian is $\hat{H}=\hat{K}+\hat{V}$, with the
perturbation (bias) given in Eq.(\ref{Vdefin}) and the modified "large" part%
\begin{equation}
\hat{K}=\int_{r}\hat{\psi}_{r}^{\dagger }H_{1Q}\hat{\psi}_{r}  \label{Kmod}
\end{equation}%
The "first quantized" operator $H_{1Q}$ now contains the barrier potential $%
U\left( r\right) $:%
\begin{equation}
H_{1Q}=-i\sigma \cdot \nabla +U\left( y\right) .  \label{H1Q}
\end{equation}%
The barrier breaks the translational symmetry, however, for the simple form
of the symmetric barrier we have adopted, Eq.(\ref{U}), the operator $H_{1Q}$
is invariant under reflection, $P:y\rightarrow -y$, \textit{supplemented by
the spinor rotation},%
\begin{equation}
S=P\sigma _{x},\text{ \ \ \ \ \ \ \ }\left[ H_{1Q},S\right] =0.  \label{S}
\end{equation}%
The bias potential, Eq.(\ref{V}), is also chosen to be antisymmetric which
simplifies the considerations.

The presence of a barrier renders the problem analogous to that in
mesoscopic physics \cite{Davies} and suggests that the T - matrix approach
to transport is useful in this case.

\begin{figure}[ptb]\begin{center}
\includegraphics[
natheight=2.0081in, natwidth=3in, height=2.0461in, width=3.0441in]
{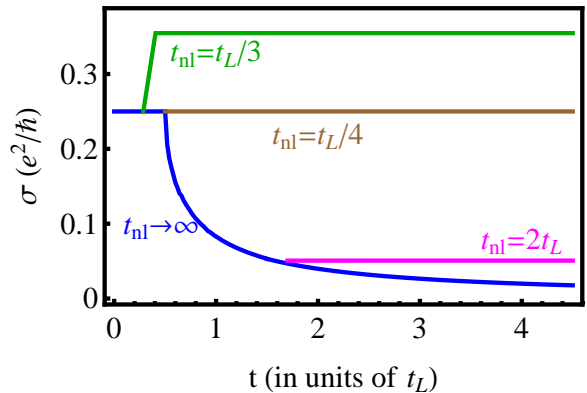}\caption{Evolution of the
scaled conductivity, Eqs.(\protect\ref{sig2},\protect\ref{sig_interpolation},%
\protect\ref{sig_inst},\protect\ref{sig_inst_large}) at Dirac point for a
finite range electric field (and no contact barrier). Electric field
strength determines the time scale $t_{nl}$, Eq.(\protect\ref{t_nl}) at
which nonlinear effects set in.}
\end{center}\end{figure}%

\subsection{B. T - matrix}

The LB\ approach utilizes the notion of a transmission coefficient through
the channel $n,$ $T_{n}\equiv \left\vert t_{n}\left( \mu \right) \right\vert
^{2}$, where $t_{n}$ is its amplitude. The conductance is 
\begin{equation}
G\left( \mu \right) =\frac{e^{2}}{h}\sum_{n}T_{n},  \label{G_LB}
\end{equation}%
where the summation is over all the open channels. Therefore one should
solve the "classical" Weyl equation with a barrier%
\begin{equation}
H_{1Q}\psi =\left[ -i\sigma \cdot \nabla +U\left( y\right) \right] \psi
=\varepsilon \psi \text{.}  \label{Weyleq}
\end{equation}%
For simplicity we consider only periodic boundary conditions in the
direction perpendicular to the field with "perimeter" $W$, although various
more realistic boundary conditions were discussed in ref. \cite{Beenakker06}
and numerous works since. Due to translational symmetry in the direction
perpendicular to the barrier, we consider a fixed value of the momentum $%
p_{x}\equiv k=\frac{2\pi }{W}n_{x}$. Despite the lack of translational
symmetry in the field direction $y$ due to the barrier, one can still use
the momentum $p_{y}\equiv p$ as a good quantum number for scattering states.
Another "number" is the sign of energy which determines the wave function in
the leads, namely distinguishes between the $u$ and the $v$ spinors given in
Eq.(\ref{uv}). The reflection symmetry defined in the previous Subsection
converts left movers into right movers%
\begin{equation}
Sv_{k,p}e^{ipy}=z_{p}v_{k,-p}e^{-ipy};\text{ \ }%
Su_{k,p}e^{ipy}=-z_{p}u_{k,-p}e^{-ipy},  \label{LR}
\end{equation}%
where we suppressed the index $k$ in $z_{kp}=\left( k+ip\right) /\varepsilon
_{\mathbf{k}}$.

The "out of barrier" equation is just the free Weyl equation with negative
and positive energy solutions, $\psi =u_{kp}e^{i\left( kx+py\right) }$
(hole) and $\psi =v_{kp}e^{i\left( kx+py\right) }$ (electron) discussed in
the previous Section. It should be matched with the "in barrier" solution.
Several distinct kinematic possibilities exist. We survey them from high to
low, see Fig. 3.

\textit{1. Energies above the barrier, }$\varepsilon >U$\textit{. }Both
inside and outside one has electron $v$- states with different momenta.
Outside the barrier $p=\sqrt{\varepsilon ^{2}-k^{2}}$, while inside the
barrier momentum in the field direction is 
\begin{equation}
q=\sqrt{\left( \varepsilon -U\right) ^{2}-k^{2}}.  \label{qdef}
\end{equation}%
One has a wave (real $q$) inside for 
\begin{equation}
p>p_{2}\equiv \sqrt{\left( U+2\left\vert k\right\vert \right) U}.
\label{1range}
\end{equation}%
The lower bound, $p_{2}\left( k\right) $, is the green line in Fig.3. There
is an evanescent (imaginary momentum $q$) particle state inside for

\begin{equation}
\sqrt{U^{2}-k^{2}}\equiv p_{U}\left( k\right) <p<p_{2}.  \label{1range1}
\end{equation}%
Crossing the red line in Fig.3, $p_{U}\left( k\right) $, one encounters
states below the barrier.

\textit{2. Positive energy states below the barrier}, $0<\varepsilon <U$.
One has the $v$ spinor (electron) outside the barrier, while the $u$ spinor
(hole) inside. For momenta $p$ in the range%
\begin{equation}
\sqrt{\left( U-2\left\vert k\right\vert \right) U}\equiv p_{1}<p<p_{U},
\label{2range}
\end{equation}%
the states are evanescent hole states. At yet lower energies, $p<p_{1},$ one
has a propagating state, but this time a \textit{hole}. This relativistic
feature is the cause of the Klein paradox.

\textit{3. Negative energy states}, $\varepsilon <0$. Outside the barrier
now one has $\varepsilon =-\sqrt{p^{2}+k^{2}}$. This is another purely
"relativistic" possibility in which one has holes both outside and hence
inside the barrier.

The Schr\"{o}dinger equation above the barrier $\varepsilon >U$ is solved by
the scattering states for right movers, $p>0$, 
\begin{eqnarray}
&&\phi _{kp}\left( y\right) =\frac{1}{\sqrt{DW}}  \label{fikp} \\
&&\times \left\{ 
\begin{array}{c}
v_{kp}e^{ipy}+r_{kp}v_{k,-p}e^{-ipy}\text{ for }y<-L/2 \\ 
A_{kp}v_{kq}e^{iqy}+B_{kp}v_{k,-q}e^{-iqy}\text{ for }-L/2<y<L/2 \\ 
t_{kp}v_{kp}e^{ipy}\text{ for }L/2<y%
\end{array}%
\right. \text{.}  \notag
\end{eqnarray}%
Matching conditions,

\begin{eqnarray}
v_{p}e^{-ipL/2}+r_{p}v_{-p}e^{ipL/2}
&=&A_{p}v_{q}e^{-iqL/2}+B_{p}v_{-q}e^{iqL/2};  \notag \\
t_{p}v_{p}e^{ipL/2} &=&A_{p}v_{q}e^{iqL/2}+B_{p}v_{-q}e^{-iqL/2},  \notag \\
&&  \label{matching}
\end{eqnarray}%
(where the fixed index $k$ is suppressed) determine the T-matrix and are
easily solved. The electron $v$ states have to be replaced with $u$ states
in the case of a hole, so in the second energy region in the barrier part $%
v\rightarrow u$, while in the third energy region in all parts $v\rightarrow
u$. For example, for evanescent modes below the barrier one obtains

\begin{equation}
t_{kp}=\frac{\left( z_{p}^{2}-1\right) \left( z_{q}^{2}-1\right) }{%
e^{-iqL}\left( 1-z_{p}z_{q}\right) ^{2}-e^{iqL}\left( z_{p}-z_{q}\right) ^{2}%
}\text{.}  \label{t_p}
\end{equation}

\begin{figure}[ptb]\begin{center}
\includegraphics[
natheight=2.9672in, natwidth=3in, height=3.0113in, width=3.0441in]
{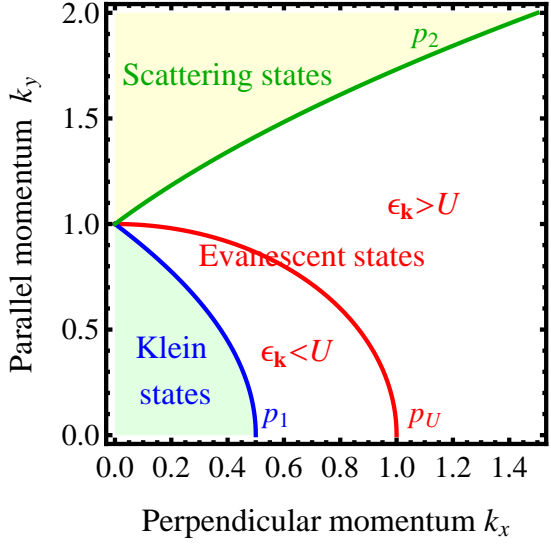}\caption{Kinematics of
scattering states of the first quantized Weyl equation with potential
barrier. The red line corresponds to $p,l=\protect\sqrt{U^{2}-k^{2}}$, the
green to $p,l=\protect\sqrt{U\left( U+2k\right) }$, and the blue to $p,l=%
\protect\sqrt{U\left( U-2k\right) }$. The lines separate kinematical regions
for the intraband transitions. Here $U=1.$}
\end{center}\end{figure}%

\subsection{C. LB conductance}

According to the LB formula one has to sum up the transmission coefficients $%
T_{kp}=\left\vert t_{kp}\right\vert ^{2}$ over all the states, namely over
the regions 1-3 that obey a constraint $p\left( k\right) =\sqrt{\mu
^{2}-k^{2}}$. The chemical potential $\mu =U+U_{gate}$ of graphene is
counted here from the bottom of the barrier and let us consider only
positive $U_{gate}$. In this case region 3, in which energy is negative, and
region 2, where energy is positive but not large enough therefore will not
contribute and we are left with region 1 of the previous Subsection:

\begin{equation}
G=\frac{4e^{2}}{2\pi }\sum_{\left\vert k\right\vert <\left\vert \mu
\right\vert }\left\vert t_{k,p\left( k\right) }\right\vert ^{2}\text{.}
\label{G}
\end{equation}%
Let us first consider, following ref.\cite{Beenakker06}, only evanescent
states contributions under the barrier. Although finite sample width with
various boundary conditions can be easily considered, we take a limit of
infinite aspect ratio $W/L\rightarrow \infty $ and generally replace
summation over $k\equiv k_{x}$ in Eq.(\ref{G}) by an integral. In this case
one can form two dimensionless combinations: the barrier "strength" $%
UL\equiv \Omega $ and $U_{gate}L$. It turns out (not shown in the present
paper) that the conductance has a smooth limit $U_{gate}L\rightarrow 0$
that, of course, involves evanescent states only. One therefore can study
directly the case of $U_{gate}=0$. Substituting the transmission coefficient
of Eq.(\ref{t_p}), one obtains the following limiting value of conductivity:

\begin{eqnarray}
\sigma _{LB}\left( \Omega \right) &=&G_{LB}\frac{L}{W}=\frac{2e^{2}LU^{2}}{%
\pi ^{2}}\int_{k=0}^{U}\frac{U^{2}-k^{2}}{U^{2}\cosh ^{2}\left( kL\right)
-k^{2}}  \notag \\
&=&\frac{2e^{2}\Omega }{\pi ^{2}}\int_{\overline{k}=0}^{1}\frac{1-\overline{k%
}^{2}}{\cosh ^{2}\left( \overline{k}\Omega \right) -\overline{k}^{2}}\text{,}
\label{G1}
\end{eqnarray}%
where the dimensionless momentum $\overline{k}=k/U$ is used. The result for
various $\Omega $ appear in Fig.4 as the long time limit.

For large $\Omega $ the integral is dominated by small $\overline{k}$ and
one gets the "mesoscopic" value of conductivity, 
\begin{equation}
\sigma _{LB}\left( \Omega \right) =\frac{2e^{2}\Omega }{\pi ^{2}}\int_{%
\overline{k}=0}^{1}\cosh ^{-2}\left( \overline{k}\Omega \right) =\frac{2e^{2}%
}{\pi ^{2}}=\sigma _{1}\text{.}  \label{sigma1}
\end{equation}%
One therefore can apply the dynamical approach to try to understand the
crossover from the short ballistic time, the electron - hole "bulk"
dynamics, to the long ballistic time, the barrier reflection dominated
dynamics.

\subsection{D. Electron tunneling into graphene from leads for $U=U_{gate}$}

The main physical effect of leads is that they "contaminate" the graphene
flake (assumed to be at Dirac point). Electrons from the lead metal tunnel
into the flake creating charge "puddles" on both sides of the barrier. As a
result the intraband channel for electric transport is greatly enhanced for
small flakes and might dominate over the interband channel, especially at
small fields. The quantitative consideration of this contamination is more
transparent in the basis of symmetric and antisymmetric eigenfunctions than
it would be in the scattering states basis (which was convenient for the LB
approach above).

The symmetry image of the scattering state $\phi _{k,p}\left( r\right) $
defined in Eq.(\ref{fikp}) is $\phi _{k,-p}\left( r\right) =\widehat{S}\phi
_{k,p}\left( r\right) $, where the symmetry operator $\widehat{S}$ was
defined in Eq.(\ref{S}). Above the barrier (region 1), $\varepsilon >U$, one
can write symmetric $\left( s\right) $/antisymmetric $\left( a\right) $
functions as (with the $x$ dependence of the wave functions $e^{ikx}$
implied and index $k$ suppressed):

\begin{eqnarray}
&&\phi _{p}^{s,a}\left( y\right) =\frac{1}{\sqrt{2}}\left( 1\pm \widehat{S}%
\right) \phi _{p}\left( y\right) =\frac{1}{\sqrt{2DW}}  \label{sym_anti1} \\
&&\times \left\{ 
\begin{array}{c}
\left( e^{ipy}+\sigma _{x}D_{p}^{s,a}e^{-ipy}\right) v_{p}\text{ for }y<-L/2
\\ 
\pm \left( e^{iqy}+\sigma _{x}e^{-iqy}\right) E_{p}^{s,a}v_{q}\text{ for }%
-L/2<y<L/2 \\ 
\pm \left( D_{p}^{s}e^{ipy}+\sigma _{x}e^{-ipy}\right) v_{p}\text{ for }L/2<y%
\end{array}%
,\right.  \notag
\end{eqnarray}%
where $q$ is the momentum in the barrier range given in Eq.(\ref{qdef}), and%
\begin{eqnarray}
D_{p}^{s,a} &=&\frac{r_{p}}{z_{p}}\pm t_{p}=e^{-ipL}\frac{z_{p}z_{q}-1\pm
\left( z_{p}-z_{q}\right) e^{-iqL}}{z_{p}-z_{q}\pm \left(
z_{p}z_{q}-1\right) e^{-iqL}};  \label{coef} \\
E_{p}^{s,a} &=&A_{p}\pm B_{p}z_{-q}=\frac{e^{-i\left( q+p\right) L/2}\left(
z_{p}^{2}-1\right) }{z_{p}-z_{q}\pm \left( z_{p}z_{q}-1\right) e^{-iqL}}%
\text{.}  \notag
\end{eqnarray}%
Below the barrier for the positive energy states (region 2) one similarly
has the same form

\begin{eqnarray}
&&\phi _{p}^{s,a}\left( y\right) =\frac{1}{\sqrt{2DW}}  \label{sym_anti2} \\
&&\times \left\{ 
\begin{array}{c}
\left( e^{ipy}+\sigma _{x}D_{p}^{\prime s,a}e^{-ipy}\right) v_{p}\text{ for }%
y<-L/2 \\ 
\left( \pm e^{iqy}+\sigma _{x}e^{-iqy}\right) E_{p}^{\prime s,a}u_{q}\text{
for }-L/2<y<L/2 \\ 
\pm \left( D_{p}^{\prime s,a}e^{ipy}+\sigma _{x}e^{-ipy}\right) v_{p}\text{
for }L/2<y%
\end{array}%
\right.  \notag
\end{eqnarray}%
with different coefficients%
\begin{eqnarray}
D_{p}^{\prime s,a} &=&-e^{-iLp}\frac{1+z_{p}z_{q}\mp \left(
z_{p}+z_{q}\right) e^{-iLq}}{z_{q}+z_{q}\mp \left( z_{p}z_{q}-1\right)
e^{-iLq}};  \label{coef2} \\
E_{p}^{\prime s,a} &=&\frac{e^{-iL\left( p+q\right) /2}\left(
z_{p}^{2}-1\right) }{z_{q}+z_{q}\mp \left( z_{p}z_{q}+1\right) e^{-iLq}};. 
\notag
\end{eqnarray}%
The negative energy states (region 3) are obtained from those in region 1 by
a replacement $v\rightarrow u$.

\begin{eqnarray}
&&\varphi _{p}^{s,a}\left( y\right) =\frac{1}{\sqrt{2DW}}  \label{sym_anti3}
\\
&&\times \left\{ 
\begin{array}{c}
\left( e^{ipy}+\sigma _{x}D_{p}^{s,a}e^{-ipy}\right) u_{p}\text{ for }y<-L/2
\\ 
\pm \left( e^{iqy}+\sigma _{x}e^{-iqy}\right) E_{p}^{s,a}u_{q}\text{ for }%
-L/2<y<L/2 \\ 
\pm \left( D_{p}^{s}e^{ipy}+\sigma _{x}e^{-ipy}\right) u_{p}\text{ for }L/2<y%
\end{array}%
\right. \text{.}  \notag
\end{eqnarray}%
We infer that physically the leads induce charge "puddles" in graphene at
Dirac point. These electrons can now be accelerated (in an ultrarelativistic
fashion by reorientation) and compete with the electron - hole channel
described above.

Now we calculate the evolution of the current density within the linear
response approximation.

\section{IV. Current evolution in graphene with barrier}

\subsection{A. The bias and the electric current in basis of the barrier
eigenstates}

The dynamics of the manybody system with the contact potential barrier after
switching on the electric field is determined in the Heisenberg picture by
the equation%
\begin{equation}
i\partial _{t}\hat{\psi}\left( t\right) =\left( K+V\right) \hat{\psi}\left(
t\right) ,  \label{Heisenberg}
\end{equation}%
where the $K$ is defined in Eq. (\ref{Kmod}) and $V$ in Eqs. (\ref{Vdefin}, %
\ref{V}).

The unperturbed part of the Hamiltonian, $\widehat{K},$ is diagonalized by
choosing a basis linked to the eigenfunctions of the first quantized
operator $H_{1Q}$, Eq.(\ref{Weyleq}):%
\begin{eqnarray}
\hat{a}_{kp}^{S} &=&\int_{r}\phi _{kp}^{S\dagger }\left( r\right) \hat{\psi}%
_{r};\hat{b}_{kp}^{S}=\int_{r}\varphi _{kp}^{S\dagger }\left( r\right) \hat{%
\psi}_{r},  \label{eigenbasis} \\
\hat{\psi}_{r} &=&\sum_{kpS}\left[ \phi _{kp}^{S}\left( r\right) \hat{a}%
_{kp}^{S}+\varphi _{kp}^{S}\left( r\right) \hat{b}_{kp}^{S}\right] ,  \notag
\end{eqnarray}%
so that$\ \hat{K}=\sum_{\lambda }\varepsilon _{kp}\left( \hat{a}_{\lambda
}^{\dagger }\hat{a}_{\lambda }-\hat{b}_{\lambda }^{\dagger }\hat{b}_{\lambda
}\right) ,$ where $\lambda $ combines $k,p,S$. The operator $\hat{a}%
^{\dagger }$ ($\hat{b}$) creates an electron (hole), and summations are over
symmetric and antisymmetric states, $S=a,s$, with $p>0$. The perturbation
operator in the new basis takes the form 
\begin{equation}
\hat{V}=\sum_{\lambda ,\sigma }\left( V_{\lambda \sigma }^{++}\hat{a}%
_{\lambda }^{\dagger }\hat{a}_{\sigma }+V_{\lambda \sigma }^{-+}\hat{b}%
_{\lambda }^{\dagger }\hat{a}_{\sigma }+V_{\lambda \sigma }^{+-}\hat{a}%
_{\lambda }^{\dagger }\hat{b}_{\sigma }+V_{\lambda \sigma }^{--}\hat{b}%
_{\lambda }^{\dagger }\hat{b}_{\sigma }\right) .  \label{Vgraphene}
\end{equation}%
For example, the electron - hole matrix element is given by 
\begin{equation}
\small%
V_{kl;k^{\prime }p}^{-+ST}=\int_{xy}e^{i\left( k^{\prime }-k\right)
x}V\left( y\right) \varphi _{kl}^{S\dagger }\left( y\right) \phi _{k^{\prime
}p}^{T}\left( y\right) =W\delta _{kk^{\prime }}\mathcal{V}_{k,lp}^{-+ST},
\label{Veh}
\end{equation}%
where 
\begin{equation}
\mathcal{V}_{k,lp}^{-+ST}=\int_{y}V\left( y\right) \varphi _{kl}^{S\dagger
}\left( y\right) \phi _{kp}^{T}\left( y\right) .  \label{Veh_def}
\end{equation}%
In the middle of the sample the current operator is

\begin{eqnarray}
&&\hat{I}_{y}\left( y=0\right) =\int_{x}\hat{\jmath}_{y}\left( x,0\right)
\label{Iy} \\
&=&\sum_{\lambda ,\sigma }\left( j_{\lambda \sigma }^{++}\hat{a}_{\lambda
}^{\dagger }\hat{a}_{\sigma }+j_{\lambda \sigma }^{-+}\hat{b}_{\lambda
}^{\dagger }\hat{a}_{\sigma }+j_{\lambda \sigma }^{+-}\hat{a}_{\lambda
}^{\dagger }\hat{b}_{\sigma }+j_{\lambda \sigma }^{--}\hat{b}_{\lambda
}^{\dagger }\hat{b}_{\sigma }\right) \text{,}  \notag
\end{eqnarray}%
where, for example,%
\begin{equation}
\text{\ }j_{k^{\prime }p;kl}^{+-TS}=W\delta _{kk^{\prime }}j_{k,pl}^{+-TS};%
\text{\ }j_{k,pl}^{+-TS}=-4e\phi _{kp}^{T\dagger }\left( 0\right) \sigma
_{y}\varphi _{kl}^{S}\left( 0\right) \text{.}  \label{jmedef}
\end{equation}%
The factor $4$ appears due to spin degeneracy and two Dirac points (valley
degeneracy). Now one can write the first order contributions to the electric
current induced by the perturbation.

\subsection{B. General expressions for the interband and the intraband
contributions to current in linear response}

In linear response one obtains at gate potential $U_{gate}$ (assumed
positive) two contributions with completely different physical
interpretations. In the first term the summation is over electron states
above $\mu =U+U_{gate}$ and electron states below the Fermi level,

\begin{eqnarray}
&&I_{ee}\left( t\right)  \label{Iee} \\
&=&-W\sum\limits_{k}\sum\limits_{p:\varepsilon _{kp}>\mu
}\sum\limits_{l:\varepsilon _{kl}<\mu }\frac{1-e^{-i\left( \varepsilon
_{kp}-\varepsilon _{kl}\right) t}}{\varepsilon _{kp}-\varepsilon _{kl}}%
\mathcal{V}_{k,lp}^{++ST}j_{k,pl}^{++TS}+cc\text{.}  \notag
\end{eqnarray}%
The summation over symmetry indices $T$ and $S$ is understood. The second
contribution sums over electrons above $\mu $ and all the hole states with
arbitrary $l$:

\begin{eqnarray}
&&I_{eh}\left( t\right)  \label{Ieh} \\
&=&-W\sum\limits_{k,l}\sum\limits_{p:\varepsilon _{kp}>\mu }\frac{%
1-e^{-i\left( \varepsilon _{kp}+\varepsilon _{kl}\right) t}}{\varepsilon
_{kp}+\varepsilon _{kl}}\mathcal{V}_{k,lp}^{-+ST}j_{k,pl}^{+-TS}+cc.  \notag
\end{eqnarray}%
The hole's momentum has no restriction since its energy is always negative,
while $U_{gate}$ is positive (in this work we do not consider the case $%
U_{gate}<0$ , that was discussed recently in ref.\cite{Sonin} in the
framework of LB approach, where other possibilities can occur). The first
contribution is the "one-particle" type (the intraband channel), very much
like in more common manybody electronic systems. The second contribution, to
the contrary, is purely ultrarelativistic (the interband channel) and
describes the electron-hole pair creation, very much like in the infinitely
long flake discussed in Sec. II.

It is useful to define an effective dimensionless conductivity. There are
only two time scales in the problem: the inverse barrier height $t_{U}=1/U$ (%
$\hbar /U$ in physical units) and the flight time to cross the sample $%
t_{L}=L$ ($L/v_{g}$) with the ratio being the only "material" parameter $%
\Omega =UL=t_{L}/t_{U}$ in the linear response considerations. For this case
the effective conductivity has a scaling property:

\begin{equation}
\sigma \left( U,L,t\right) =\frac{I\left( U,L,t\right) }{V_{0}}\frac{L}{W}%
=\sigma \left( \Omega ,\overline{t}\right) .  \label{sigma_scaled}
\end{equation}%
The scaled time is defined, as in Sec. II, $\overline{t}=t/t_{L}$. Since in
the dynamical approach the limit $U_{gate}\rightarrow 0$ is smooth \cite%
{Lewkowicz09}, in what follows the conductivity will be calculated directly
for $U_{gate}=0.$We concentrate on the case when quantizations of $k_{x}$
and $k_{y}$ are not important (the generalization to the discrete case is
straight forward), the sums in Eqs. (\ref{Iee}) and (\ref{Ieh}) can be
replaced by integrals over dimensionless momentum variables, $k=U\overline{k}
$, $p=U\overline{p}$ and $\overline{p}_{U}=\sqrt{1-\overline{k}^{2}}$. The
intraband contribution becomes 
\begin{eqnarray}
&&\sigma ^{ee}\left( \overline{t}\right) =-2\frac{LU^{2}}{V_{0}}\frac{W}{%
2\pi }\left( \frac{D}{2\pi }\right) ^{2}  \label{sigee} \\
&&\times \int_{\overline{k}=0}^{1}\int_{\overline{l}=0}^{\overline{p}%
_{U}}\int_{\overline{p}=\overline{p}_{U}}^{\infty }\frac{1-e^{-i\left(
\epsilon -\epsilon ^{\prime }\right) \Omega \overline{t}}}{\epsilon
-\epsilon ^{\prime }}\mathcal{V}_{lp}^{++ST}j_{pl}^{++TS}+cc\text{.}  \notag
\end{eqnarray}%
Here rescaled energies, $\epsilon \equiv \varepsilon _{kp}/U,$ $\epsilon
^{\prime }\equiv \varepsilon _{kl}/U$ in addition to momenta $\overline{k},%
\overline{p},\overline{l}$ are used and the factor $2$ comes from positive
and negative $k$. Similarly the interband contribution to conductivity takes
the form%
\begin{eqnarray}
&&\sigma ^{eh}\left( \overline{t}\right) =-2\frac{LU^{2}}{V_{0}}\frac{W}{%
2\pi }\left( \frac{D}{2\pi }\right) ^{2}  \label{sig_eh} \\
&&\times \int_{\overline{k}=0}^{1}\int_{\overline{l}=0}^{\infty }\int_{%
\overline{p}=\overline{p}_{U}}^{\infty }\frac{1-e^{-i\left( \epsilon
+\epsilon ^{\prime }\right) \Omega \overline{t}}}{\epsilon +\epsilon
^{\prime }}\mathcal{V}_{lp}^{-+ST}j_{pl}^{+-TS}+cc\text{.}  \notag
\end{eqnarray}

Next we consider the two contributions in turn.

\section{V. The intraband contribution}

\subsection{A. The electron - electron matrix elements in linear response}

Neglecting the dependencies on the cutoff and keeping notations used in Sec.
IV A one obtains, after some algebra, the following expressions for the
matrix elements for the electron - electron contribution for $T\neq S$,%
\begin{eqnarray}
&&\mathcal{V}_{lp}^{++TS}=\frac{ieV_{0}}{2DW}  \label{Vmeee} \\
&&\times \left\{ \left( 1+z_{l}^{\ast }z_{p}\right) \frac{D_{l}^{\prime
T\ast }D_{p}^{S}e^{i\left( p-l\right) L/2}-e^{-i\left( p-l\right) L/2}}{p-l}%
\right.  \notag \\
&&+\left( z_{p}+z_{l}^{\ast }\right) \frac{D_{p}^{S}e^{i\left( p+l\right)
L/2}-D_{l}^{\prime T\ast }e^{-i\left( p+l\right) L/2}}{p+l}  \notag \\
&&\left. -4E_{l}^{^{\prime }T\ast }E_{p}^{S}\left[ \left( 1-z_{q}z_{g}^{\ast
}\right) f\left( g-q\right) +\left( z_{q}-z_{g}^{\ast }\right) f\left(
g+q\right) \right] \right\} .  \notag
\end{eqnarray}%
where the function $f$ is defined as 
\begin{equation}
f(q)=\frac{\sin \left( qL/2\right) }{q^{2}}-\frac{L}{2}\frac{\cos \left(
qL/2\right) }{q}.  \label{function}
\end{equation}

It is important to note that the dominant contributions to the integrals in
Eq.(\ref{sigee}) come from the $\left( p-l\right) ^{-1}$ poles. The
corresponding current density matrix elements are:%
\begin{eqnarray}
j_{pl}^{++sa} &=&i\frac{e}{WD}E_{p}^{s\ast }E_{l}^{\prime a}\left(
1+z_{q}^{\ast }\right) \left( 1+z_{g}\right) ;  \label{jmeee} \\
j_{pl}^{++as} &=&-i\frac{e}{WD}E_{p}^{a\ast }E_{l}^{\prime s}\left(
1-z_{q}^{\ast }\right) \left( 1-z_{g}\right) .  \notag
\end{eqnarray}

The one-particle (electron - electron) contribution to the scaled
conductivity, Eq.(\ref{sigma_scaled}) is shown in Fig. 4 as red curves for
various values of $\Omega $ as function of time $\overline{t}$. At times
shorter than both $t_{L}$ and $t_{U}$ it rises linearly. For $\Omega >1$ the
conductivity oscillates before eventually approaching the LB result of Sec.
III C. In the following subsections the two ballistic regimes, namely, the
limit of long times $t>>t_{U},t_{L}$ (or in scaled form $\overline{t}%
>>1/\Omega ,1$) and short times $t<<t_{U},t_{L}$ (or in scaled form $%
\overline{t}<<1/\Omega ,1$) are analyzed analytically.

\subsection{B. Short time asymptotics}

Expanding the electron - electron contribution to the conductivity, Eq.(\ref%
{sigee}), to leading order in the short time limit yields

\begin{equation}
\sigma ^{ee}=-i\overline{t}\frac{e^{2}\Omega }{\pi ^{3}}\int_{\overline{k}%
=0}^{1}\int_{\overline{l}=0}^{\overline{p}_{U}}\int_{\overline{p}=\overline{p%
}_{U}}^{\infty }\mathcal{V}_{lp}^{++ST}j_{pl}^{++TS}+cc\text{.}  \label{B1}
\end{equation}%
This leads to a small conductivity contribution since the integration range
on the variable $\overline{l}$ is rather limited. The linear behaviour of
the conductivity is shown in Fig. 4 (red lines). \ For the case $\Omega
\lesssim 1$ ($t_{L}<t_{U}$), represented in Fig.4 by $\Omega =$ $\pi /16$, $%
\pi /4$, the intraband contribution is positive and increases monotonically.
However, when $\Omega >1$ ($t_{L}>t_{U}$) represented in Fig.4 by $\Omega =$ 
$\pi $, $2\pi $, it becomes negative.

\subsection{C. Long time asymptotics}

Due to the oscillating functions in Eq.(\ref{sigee}) the long time
asymptotics is due solely to the region of the three dimensional integral
when $\varepsilon _{kp}-\varepsilon _{kl}\rightarrow 0$. Consequently, in
view of the discussion of the various kinematical regions in Sec. III B,
summarized in Fig.3, the limit is dominated by\ integrating over the
transitions from evanescent states above the barrier (region 1) to
evanescent states below the barrier (region 2). Since the integral is
dominated by the neighbourhood of the line $\varepsilon _{kp},\varepsilon
_{kl}\approx U$, one can expand around this line (the red line in Fig.3), $%
\overline{p}\approx \overline{p}_{U}+\Delta p$, so that the evanescent
momentum (in the field direction) is, to second order, $\overline{k}-\frac{%
\overline{p}_{U}^{2}}{2\overline{k}}\Delta p^{2}$. Similar expansions are
made for the momentum $\overline{l}$. Therefore one can simplify the
expression for the conductivity by replacing the integrand in Eq.(\ref{sigee}%
) by its limiting value. Albeit, this limit is nontrivial. In particular,
the lower components of the spinor in Eq.(\ref{uv}) is just 
\begin{equation}
z_{p}\rightarrow Z\equiv \overline{k}+i\overline{p}_{U},\text{ \ }z_{q}=%
\frac{\overline{p}_{U}}{2\overline{k}}\Delta p\text{.}  \label{zlimit}
\end{equation}%
As a result the coefficients determining the eigenstates, Eqs.(\ref{coef},%
\ref{coef2}), simplify considerably:

\begin{eqnarray}
E_{p}^{a,s} &=&E_{l}^{\prime a,s}\rightarrow e^{\Omega \left( \overline{k}-i%
\overline{p}_{U}\right) /2}\frac{Z^{2}-1}{Z\pm e^{\Omega \overline{k}}};%
\text{ \ }  \label{coeflimit} \\
\text{\ }D_{p}^{a,s} &=&D_{l}^{\prime a,s}\rightarrow -e^{-i\Omega \overline{%
p}_{U}}\frac{1+Ze^{\Omega \overline{k}}}{Z\pm e^{\Omega \overline{k}}}\text{.%
}  \notag
\end{eqnarray}

Since at small $\kappa $, $f\left( \kappa \right) \simeq \frac{L^{3}}{24}%
\kappa $, the third contribution to the perturbation matrix elements, Eq.(%
\ref{Vmeee}), is negligible. The results for the matrix elements of the
perturbation and the current are

\begin{eqnarray}
\mathcal{V}_{lp}^{as} &=&-\mathcal{V}_{lp}^{sa\ast }=\frac{2eVU}{DW}\frac{%
\overline{p}_{U}}{i\overline{p}_{U}-\sinh \left( \Omega \overline{k}\right) }%
\frac{1}{\overline{p}-\overline{l}};  \label{VJlimit} \\
j_{pl}^{sa} &=&j_{pl}^{as\ast }=-\frac{8ie}{DW}\frac{\overline{p}_{U}^{2}}{i%
\overline{p}_{U}+\sinh \left( \Omega \overline{k}\right) }.  \notag
\end{eqnarray}%
Substituting these matrix elements into Eq.(\ref{sigee}), one obtains the
contribution to conductance:

\begin{eqnarray}
\sigma ^{ee}\left( \overline{t}\right) &=&\frac{16e^{2}\Omega }{\pi ^{3}}%
\int_{\overline{k}=0}^{1}\frac{\overline{p}_{U}^{3}\left( \overline{k}%
\right) }{\cosh \left( \Omega \overline{k}\right) ^{2}-\overline{k}^{2}}
\label{cond_limit} \\
&&\times \int_{\overline{l}=0}^{\overline{p}_{U}}\int_{\overline{p}=%
\overline{p}_{U}}^{\infty }\frac{\sin \left[ \left( \epsilon -\epsilon
^{\prime }\right) \Omega \overline{t}\right] }{\epsilon -\epsilon }\frac{1}{%
\overline{p}-\overline{l}}\text{,}  \notag
\end{eqnarray}

Since $\overline{p}-\overline{l}\approx \left( \epsilon -\epsilon ^{\prime
}\right) /\overline{p}_{U}$, one gets for the last two integrals

\begin{eqnarray}
&&\overline{p}_{U}\left( k\right) \int_{\epsilon ^{\prime }=\overline{k}%
}^{1}\int_{\epsilon =1}^{\infty }\frac{\epsilon \epsilon ^{\prime }}{\sqrt{%
\epsilon ^{2}-\overline{k}^{2}}\sqrt{\epsilon ^{\prime 2}-\overline{k}^{2}}}%
\frac{\sin \left[ \left( \epsilon -\epsilon ^{\prime }\right) \Omega 
\overline{t}\right] }{\left( \epsilon -\epsilon ^{\prime }\right) ^{2}} 
\notag \\
&\approx &\frac{1}{\overline{p}_{U}\left( k\right) }s\left( \overline{k}%
,\Omega \overline{t}\right) ,  \label{sigmat}
\end{eqnarray}%
where the function 
\begin{equation}
s\left( \overline{k},\tau \right) =\int_{\epsilon ^{\prime }=\overline{k}%
}^{1}\int_{\epsilon =1}^{\infty }\frac{\sin \left[ \left( \epsilon -\epsilon
^{\prime }\right) \tau \right] }{\left( \epsilon -\epsilon ^{\prime }\right)
^{2}}\rightarrow _{\tau \rightarrow \infty }\frac{\pi }{2}\text{.}  \label{s}
\end{equation}%
\qquad Therefore at large times $t>t_{U}$, one finally obtains%
\begin{equation}
\sigma ^{ee}\left( \Omega \right) =\frac{2e^{2}\Omega }{\pi ^{2}}\int_{%
\overline{k}=0}^{1}\frac{1-\overline{k}^{2}}{\cosh ^{2}\left( \Omega 
\overline{k}\right) -\overline{k}^{2}}=\sigma _{LB}\left( \Omega \right) 
\text{.}  \label{Gfinal}
\end{equation}

This is one of the main results of the paper. The electron - electron
contribution converges to the LB result, Eq.(\ref{G1}) at large times.
Furthermore, with large $\Omega $ the conductivity is dominated by
infinitesimal $\overline{k}$ and consequently can be written as

\begin{equation}
\sigma ^{ee}=\frac{e^{2}\Omega }{\pi ^{3}}\frac{\pi }{2}\int_{\overline{k}%
=0}^{1}\frac{1}{\cosh ^{2}\left( \overline{k}\Omega \right) }=\frac{e^{2}}{%
2\pi }\frac{1}{\pi }\text{.}  \label{sigma_limit}
\end{equation}%
Thus, for four flavours, we reproduce $\sigma _{1},$ Eq.(\ref{sig1}),
starting from the dynamical approach .

\section{VI. The interband contribution}

\subsection{A. The electron - hole matrix elements in linear response}

Similarly to the above, the matrix elements for the electron - hole
Landau-Zener-Schwinger contribution read (with $T\neq S$):

\begin{eqnarray}
&&\mathcal{V}_{lp}^{-+TS}=\frac{ieV}{2DW}  \label{Vmeh} \\
&&\times \left\{ \left( z_{p}-z_{l}^{-1}\right) \frac{D_{p}^{S}e^{i\left(
p+l\right) L/2}+D_{l}^{\ast S}e^{-i\left( p+l\right) L/2}}{p+l}\right. 
\notag \\
&&-\left( 1-z_{p}z_{l}^{-1}\right) \frac{D_{l}^{\ast S}D_{p}^{S}e^{i\left(
p-l\right) L/2}+e^{-i\left( p-l\right) L/2}}{p-l}  \notag \\
&&\left. +4E_{l}^{S\ast }E_{p}^{S}\left[ \left( 1-z_{q}z_{g}^{-1}\right)
f\left( g-q\right) +\left( z_{q}-z_{g}^{-1}\right) f\left( g+q\right) \right]
\right\} \text{.}  \notag
\end{eqnarray}%
Unlike the intraband contribution of the previous Section the $p-l$ pole is
cancelled by a numerator (both $p$ and $l$ are positive) and therefore the
contribution will not be dominated by the evanescent states. The
correspondent current density matrix elements are

\begin{eqnarray}
j_{pl}^{+-sa} &=&-\frac{ie}{WD}E_{p}^{s\ast }E_{l}^{s}\left(
1+z_{q}^{-1}\right) \left( 1+z_{g}\right)  \label{jmeeh} \\
j_{pl}^{+-as} &=&\frac{ie}{WD}E_{p}^{a\ast }E_{g}^{a}\left(
1-z_{q}^{-1}\right) .  \notag
\end{eqnarray}

The expression for conductivity $\sigma ^{eh}\left( U,L,t\right) $ is UV
divergent like the conductivity of the infinite sample $\sigma \left(
U=0,L,t\right) \equiv \sigma _{0}\left( \overline{t}\right) $ biased in the
region of length $L$ that was studied in Sec. II. B (and which is solely due
to the electron - hole pairs). The difference $\Delta \sigma ^{eh}\left( 
\overline{t}\right) =\sigma ^{eh}\left( U,L,t\right) -\sigma _{0}\left( 
\overline{t}\right) $ however is finite. Using Eq.(\ref{sig_eh}) for $\sigma
^{eh}$, the difference can be written as an integral over a limited domain
of momenta in the field direction:

\begin{eqnarray}
&&\Delta \sigma ^{eh}\left( \overline{t}\right) =-\frac{\Omega U\left(
WD\right) ^{2}}{4\pi ^{3}V_{0}}  \label{delta_sigma} \\
&&\times \int_{\overline{l}=0}^{\infty }\int_{\overline{k}=0}^{1}\int_{%
\overline{p}=0}^{\overline{p}_{U}}\frac{1-e^{-i\left( \epsilon +\epsilon
^{\prime }\right) \Omega \overline{t}}}{\epsilon +\epsilon ^{\prime }}%
\mathcal{V}_{lp}^{++ST}j_{pl}^{++TS}+cc.  \notag
\end{eqnarray}%
The results, for $U_{gate}=0,$ are given in Fig.4 for several $\Omega $ as
blue lines and were discussed in Sec. II. At small times it starts with the
ultrarelativistic value $\sigma _{2}=\frac{1}{4}$ and at relatively large
times ($t_{U}>>t>t_{L}$) it decays as $\frac{1}{4\pi \overline{t}}$. \ This
short time value $\sigma _{2}$ does not change when $U>0$ provided the time
is smaller than $t_{L}/2$. This follows from the fact that in relativistic
graphene information about barrier cannot arrive at the center of the sample
before that time. The long time asymptotics in the presence of the barrier
is considered next.

\subsection{B. Long time asymptotics}

Due to oscillations the long time behaviour of the electron - hole
contribution is dominated by the region $\varepsilon _{kp}+\varepsilon
_{kl}\rightarrow 0$ in the integrations in Eq.(\ref{delta_sigma}). Unlike
for the intraband contribution in the last Section, in this case all three
momenta $k$,$p$,$l$ \ in Eq.(\ref{delta_sigma}) are small. In this limit $%
q,g\rightarrow 1,$ and hence the phases $z_{q},z_{g}\rightarrow i$ and the
function $f\left( g+q\right) \rightarrow -\frac{1}{4}\left[ \Omega \cos
\left( \Omega \right) -\sin \left( \Omega \right) \right] $ and $f\left(
g-q\right) \rightarrow 0$. Therefore it is simple to calculate $\Delta
\sigma ^{eh}$ for special values of $\Omega $. This is done in Appendix.
Numerical results show that $\Delta \sigma ^{eh}\left( \Omega =3\pi /2+2\pi
N\right) $, $\Delta \sigma ^{eh}\left( \pi /2+2\pi N\right) \propto 1/%
\overline{t}^{3}$, while $\Delta \sigma ^{eh}\left( \pi +2\pi N\right)
=-\Delta \sigma ^{eh}\left( 2\pi N\right) \approx 1/4\pi \overline{t}$. One
can fit the long time asymptotics as $\Delta \sigma ^{eh}\symbol{126}\cos
\left( \Omega \right) /4\pi \overline{t}$.

\begin{figure}[ptb]\begin{center}
\includegraphics[
natheight=3.8821in, natwidth=5.5564in, height=2.437in, width=3.4411in]
{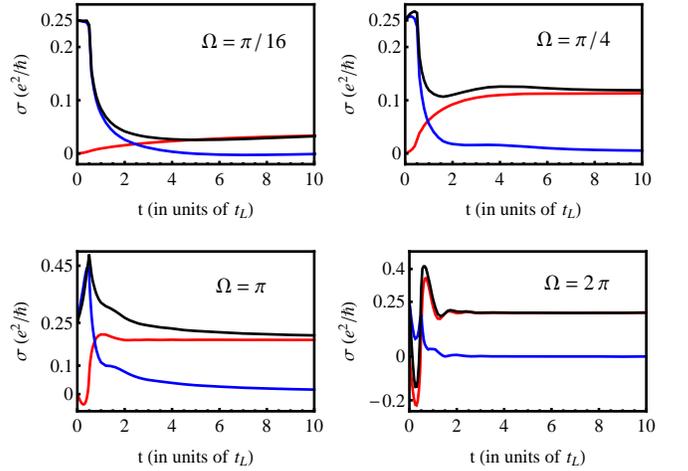}\caption{Conductivity of
finite length samples for $\Omega =UL/\hbar v_{g}=\protect\pi /16,\protect%
\pi /4,\protect\pi ,2\protect\pi $. The intraband contribution is the red
line, interband contribution - blue lines. The total conductivity - the
black lines. The time is given in units of $t_{L}=L/v_{g}$, while unit of
conductivity is $e^{2}/\hbar $.}
\end{center}\end{figure}%

\section{ VII. Discussion and conclusions}

Two different kinds of ballistic behaviour occur in undoped graphene
(graphene at Dirac point) at zero temperature. The first one is a very
unusual "ultra - relativistic" interband physics. Electron - hole pairs are
copiously created via Landau-Zener-Schwinger's mechanism by an applied
electric field. It is not dependent on leads and finite size effects of the
graphene sample. On the contrary, the second one, the intraband physics is
mostly sensitive to finite size effects and contacts. The interband
transition results, within linear response, in the universal bulk value $%
\sigma _{2}$ of conductivity, while the intraband transition is
characterized by a shape dependent linear response with the effective
conductivity $\sigma _{1}$ for large aspect ratio rectangular flakes.

Now we recapitulate under what conditions either of these two processes is
dominant in experiments on a time scale $1/\omega $ in an AC electric field $%
E$ (or in a pulse of duration $1/\omega $) for a graphene flake of length $L$
(in the limit of large width $W,$ although the discussion can be easily
extended to finite aspect ratios) and a contact barrier potential $U$. Here
we classify various practically important ranges of sample ($U,L$) and
experimental $\left( \omega ,E\right) $ parameters.

\subsection{A. "Unintrusive" experiments, $U=0$}

In reflectance and transmission experiments in visible to mid IR or even
microwave range \cite{GeimScience08} there are no leads, hence no potential
barrier, $U=0$. The interband (Landau - Zener - Schwinger) process is
dominant for any practically length and electric field $E$. However, the
transport can be either linear or highly nonlinear.

(i) For $1/\omega <t_{nl}=\sqrt{\hbar /eEv_{g}}$ one has linear response, $%
J=\sigma _{2}E$, with the interband value of conductivity $\sigma _{2}=\frac{%
\pi }{2}\frac{e^{2}}{h}$. In this case the nonlinear Schwinger's pair
creation regime is not yet reached. The conductivity is real and frequency
independent (no inductive part). There are small frequency dependent
corrections to linear response (the $E^{3}$ correction to both real and
imaginary part and third harmonic generation was calculated in ref.\cite%
{Kao10}, see Eqs.(53), (54) therein).

(ii) For $1/\omega >t_{nl}$ \textit{and} $t_{nl}<t_{L}=L/v_{g}$ the
transport is still dominated by electron - hole \ channel, but is nonlinear.
The electron - hole pairs are efficiently created due to the LZS mechanism
with rate proportional to $E^{3/2}$. This results in following I - V curve,
see Eq.(\ref{sig_inst_large}):

\begin{equation}
J=\frac{eL}{\pi ^{2}}\left( \frac{eE}{\hbar }\right) ^{3/2}\text{.}
\label{IV}
\end{equation}

(iii) For $1/\omega >t_{nl}$ \textit{and} $t_{nl}>t_{L}$ the transport is
still dominated by electron - hole \ channel and the LZS process but since
the electric field is applied in the limited space (length $L$ which is not
large enough) and the current is much smaller, see Eq.(\ref{sig_inst_small}):

\begin{equation}
J=\frac{2eL^{2}}{\pi ^{2}v_{g}}\left( \frac{eE}{\hbar }\right) ^{2}\text{.}
\label{IV_small}
\end{equation}

\subsection{B. Large barrier}

In samples on substrate with metallic leads the work function of the
graphene and the metal is typically different and as a result the contact
potential difference is of order $U=0.1-1eV$, see calculations in \cite%
{Khomyakov10} and references therein. It can be both positive and negative.
In this case the corresponding time scale $t_{U}=\hbar /U<7fs$ and typically
smaller than any of the other scales $t_{L}=L/v_{g},t_{nl}=\sqrt{\hbar
/eEv_{g}}$. This leads to an effective suppression of the electron - hole
channel for all the frequencies in the infrared range and smaller (including
DC) and the physics is dominated by the electron - electron channel.

(i) $1/\omega >t_{U}$. The DC conductance is given by the Landauer - B\"{u}%
tticker formula, generalized in Eq.(\ref{G1}), and is more sensitive to the
properties of the leads than those of graphene. When graphene is "nominally"
at Dirac point, namely, when the chemical potential of the lead is on the
barrier, graphene is still contaminated by charges tunneling into the stripe
from the leads. These electrons are accelerated and lead to the mesoscopic
type of conductance. For a large aspect ratio the effective conductivity is $%
\sigma _{1}=\frac{4}{\pi }\frac{e^{2}}{h}$. \ The assumption of an
"infinite" barrier was made early on in \cite{Beenakker06} in order to
develop the mesoscopic approach to transport in graphene.

(ii) $1/\omega <t_{U}$. The high frequencies (microwave and above)
experiments are done without leads. However if one had a set-up with leads
it could not significantly alter the pseudo-Ohmic behaviour with $\sigma
=\sigma _{2}$ since the contaminated regions constitute only a small
fraction of the sample. There is no effect of ballistic acceleration across
the sample for large frequencies or short pulses.

\subsection{C. Small barrier}

For certain materials and geometries the potential barrier may be much
smaller, of the order $0.01eV$ (for example the $Ti$ lead in the $\beta =\pi
/2$ geometry has $-0.04$, while $Ni$ in geometry $\beta =\pi $ has $0.04eV,$
see reference \cite{Khomyakov10} for details). It is feasible that the
barrier even can be tuned to zero. It is not clear what is the contact
barrier in the suspended graphene systems. It is reasonable to assume that
for suspended samples the barrier is smaller due weaker coupling. Moreover,
even in two probe experiments on suspended graphene the Dirac point appeared
exactly at zero bias voltage\cite{AndreiNN08,BolotinPRL08} (compared to up
to $40V$ in sampled on substrate) and there are less charge "puddles". That
also signals that \ the leads do not create the puddles. In such a case a
variety of crossover phenomena might occur when $t_{U}$ is comparable to $%
t_{nl}$ or $t_{L}$ studied here.

\subsection{D. Finite size and topology effects}

The calculations made in the present paper can be easily generalized to a
finite or even small flake width $W$ (assumed infinite here). In fact, it
was shown numerically\cite{Cini10}, that in graphene nanoribbons modeled by
a tight binding model one observes a crossover from $\sigma _{2}$ to $\sigma
_{1ribbon}=\frac{e^{2}}{h}$ at a time roughly corresponding to $t_{L}$. The
value of $\sigma _{1ribbon}\,$\ in this model is different from $\frac{4}{%
\pi }\frac{e^{2}}{h}$ due to the different topology and was calculated in 
\cite{OnipkoPRB08}. In Fig. 3 of \cite{Cini10} the time evolution of current
for $t=0-10^{3}t_{L}$ is shown. The current rises from zero and settles on $%
\sigma _{2}$(via oscillations) on the microscopic time scale of $t_{\gamma
}=\hbar /\gamma \approx 2.6\cdot 10^{-16}s.$ This time scale is not seen in
the Weyl model (it appears as an ultraviolet cutoff). Yet similar results,
including oscillations, were obtained in the solution of the tight binding
model of the infinite sample\cite{Lewkowicz09}. The crossover from the
initial interband to the intraband behaviour also occurs as a series of
oscillations around the value $\sigma _{1ribbon}$. The period of
oscillations in Fig.3 in reference \cite{Cini10} is roughly $t_{L}$, as it
is also the case in our Fig.4 for large $\Omega $. It is not easy to compare
these results directly, since the microscopic model for leads cannot be
readily translated into a potential barrier $U$ of the continuum model.
However, the similarity with our results suggests that $\Omega =UL/\hbar
v_{g}>>1,$ since their number of oscillations is large.

Our calculation is trivially extended to any system with Dirac point - like
spectrum like double layer graphene and recently synthesized family of
materials called "topological insulators" \cite{topins} in which a surface
excitations are similar to those in graphene with notable exception of the
chirality. Schwinger's mechanism is also expected in these materials since
chirality was not involved (left and right movers contributed equally to the
emission rate of graphene). On the contrary, the effects of the sample
topology probably require applications of sophisticated methods like
conformal mapping used in \cite{Rycerz09}.

Finally, let us remark on the role of the Klein paradox in ballistic
transport in graphene. While the pair creating mechanism can be linked to
the Klein paradox behaviour, the intraband mechanism at Dirac point \textit{%
does not make use} of the states undergoing the Klein paradox. The Klein
tunneling states described in Sec. III B (2) are characterized by energies
in the range $0<\varepsilon <U,$ the states below the blue line in Fig. 3.
These states make a contribution to the current at finite times, however
their contribution to current at large times vanishes.

\textbf{Acknowledgements}. We are indebted to K.H. Wu, E. Farber and W. B.
Jian, V. Nazarov, H.C. Kao, Y. Lin and S. Li for valuable discussions. Work
of B.R. and D.N. was supported by NSC of R.O.C. Grants No.
98-2112-M-009-014-MY3, the National Center for Theoretical Sciences, and MOE
ATU program. M.L. acknowledges the hospitality and support of the NCTU,
Electrophysics Dept and MOE ATU program.

\bigskip

\section{Appendix. Long time asymptotics of the interband contribution for
integer values of $\Omega /\left( 2\protect\pi \right) $}

In the long time limit of the e-h conductivity, Eq.(\ref{delta_sigma}), the
"under the barrier' momenta have a simple limit: $q,g\rightarrow 1,$ so that
the phases $z_{q},z_{g}\rightarrow i$ and the function\ $f\left( g+q\right)
\rightarrow -\frac{1}{4}\left[ \Omega \cos \left( \Omega \right) -\sin
\left( \Omega \right) \right] $ and $f\left( g-q\right) \rightarrow 0$.

When $\Omega =2n\pi ,$ the wave function coefficients in Eq.(\ref{coef})
reduce to%
\begin{eqnarray}
E_{p}^{s} &=&\frac{\left( -1\right) ^{n}\left( 1-i\right) }{2}\left(
z_{p}+1\right) ;  \label{Ap1} \\
E_{p}^{a} &=&\frac{\left( -1\right) ^{n}\left( 1+i\right) }{2}\left(
z_{p}-1\right)  \notag \\
D_{p}^{s} &=&-D_{p}^{a}=1.  \notag
\end{eqnarray}%
Similarly for $\Omega =\left( 2n+1\right) \pi ,$the wave function
coefficients in Eq.(\ref{coef}) become%
\begin{eqnarray}
E_{p}^{s} &=&\frac{\left( -1\right) ^{n}\left( 1-i\right) }{2}\left(
z_{p}-1\right) ;  \label{Ap2} \\
E_{p}^{a} &=&-\frac{\left( -1\right) ^{n}\left( 1+i\right) }{2}\left(
z_{p}+1\right) ;  \notag \\
D_{p}^{a} &=&-D_{p}^{s}=1.  \notag
\end{eqnarray}%
Thus the difference in conductivity due to barrier, Eq.(\ref{delta_sigma}),
when $\Omega $ is even or odd integer of $\pi ,$ will clearly depend on $n$
as (see notations in text)%
\begin{eqnarray}
&&\Delta \sigma ^{eh}\left( \Omega \right)  \label{Ap3} \\
&=&\mp \frac{2\Omega e^{2}}{\pi ^{3}}\int_{\overline{l}=0}^{\infty }\int_{%
\overline{k}=0}^{1}\int_{_{\overline{p}=0}}^{\overline{p}_{U}}\frac{\sin %
\left[ \left( \epsilon +\epsilon ^{\prime }\right) \Omega \bar{t}\right] }{%
\epsilon +\epsilon ^{\prime }}  \notag \\
&&\times \left[ \frac{1}{\overline{p}+\overline{l}}\left( \frac{\overline{p}%
}{\epsilon }+\frac{\overline{l}}{\epsilon ^{\prime }}\right) +\frac{1}{%
\overline{p}-\overline{l}}\left( \frac{\overline{p}}{\epsilon }-\frac{%
\overline{l}}{\epsilon ^{\prime }}\right) -1-\frac{\overline{k}^{2}}{%
\epsilon \epsilon ^{\prime }}\right] ,  \notag
\end{eqnarray}%
where "$-$" and "$+$" are corresponding to even and odd integer of $\Omega
/2\pi $, respectively. Numerical results show that $\Delta \sigma
^{eh}\left( \Omega =2n\pi \right) \approx -\Delta \sigma ^{eh}\left[ \Omega
=\left( 2n+1\right) \pi \right] \approx \frac{1}{4\pi \overline{t}}$.

When $\Omega =\left( 2n\pm 1/2\right) \pi $, wave\ function coefficients in
Eq.(\ref{coef}) can be rewritten as 
\begin{eqnarray}
E_{p}^{s} &=&\frac{\left( -1\right) ^{n}e^{-i\pi /4}}{2}z_{p}^{-1}\left(
z_{p}^{2}-1\right)  \label{Ap4} \\
E_{p}^{a} &=&\frac{i\left( -1\right) ^{n}e^{-i\pi /4}}{2}\left(
z_{p}^{2}-1\right)  \notag \\
D_{p}^{s} &=&-z_{p}^{-1};D_{p}^{a}=-z_{p}  \notag
\end{eqnarray}%
and 
\begin{eqnarray}
E_{p}^{s} &=&\frac{i\left( -1\right) ^{n}e^{i\pi /4}}{2}\left(
z_{p}^{2}-1\right) ;  \label{Ap5} \\
E_{p}^{a} &=&\frac{\left( -1\right) ^{n}e^{i\pi /4}}{2}z_{p}^{-1}\left(
z_{p}^{2}-1\right) ;  \notag \\
D_{p}^{s} &\rightarrow &-z_{p};D_{p}^{a}=-z_{p}^{-1};  \notag
\end{eqnarray}%
respectively. Hence one of products of matrix elements vanish $\mathcal{V}%
_{lp}^{-+as}j_{pl}^{+-sa}+\mathcal{V}_{lp}^{-+sa}j_{pl}^{+-as}=0$. That
renders the difference of conductivity independent on $\Omega :$%
\begin{equation}
\Delta \sigma ^{eh}=\mp 2\frac{e^{2}}{\pi ^{3}}\int_{\overline{l}=0}^{\infty
}\int_{\overline{k}=0}^{1}\int_{\overline{p}=0}^{\overline{p}_{U}}\frac{\sin %
\left[ \left( \epsilon +\epsilon ^{\prime }\right) \Omega \overline{t}\right]
}{\epsilon +\epsilon ^{\prime }}\frac{\overline{p}^{2}\overline{l}^{2}}{%
\epsilon ^{2}\epsilon ^{\prime 2}}\sim \overline{t}^{-3}.  \label{Ap6}
\end{equation}

\end{document}